\documentclass[11pt,journal]{IEEEtran}
\usepackage[utf8]{inputenc}
\usepackage{cite}
\usepackage{float}
\usepackage{graphicx}
\usepackage[frozencache, cachedir=.]{minted}
\usepackage{listings}
\usepackage{amsmath}
\usepackage{multirow}
\usepackage{algorithm}
\usepackage{algpseudocode}
\usepackage{hyperref}
\usepackage[none]{hyphenat}
\usepackage[margin=2cm]{geometry}

\begin{document}

\title{\textbf{Simulation of Video Streaming Over Wireless Networks with NS-3}}
\author{\IEEEauthorblockN{Guoxi Liu\IEEEauthorrefmark{1} and Liren Kong\IEEEauthorrefmark{2}\\}
\IEEEauthorblockA{School of Computing, Clemson University, Clemson, SC, USA.\\
E-mail: \IEEEauthorrefmark{1}guoxil@clemson.edu,
\IEEEauthorrefmark{2}lirenk@clemson.edu}}

\IEEEtitleabstractindextext{%
\begin{abstract}
    The combination of video streaming services and wireless networks plays an important role in many fields. In this paper, we present an NS-3-based simulation platform for evaluating and optimizing the performance of the video streaming application over wireless networks. The simulation is designed to provide real-time measurements, and thus it saves the high costs for real equipment. The developed platform consists of the video streaming server, video streaming client, and wireless network environment, where the video streaming operation is implemented under UDP protocol and is equipped with an application-level adaptive rate controller. We use the platform to simulate different use cases of sharing a wireless multi-access channel among multiple video streams.
\end{abstract}

}

\maketitle

\IEEEdisplaynontitleabstractindextext

\section{Introduction}
\label{sec:introduction}

Wireless networks are very common in modern life, which are deployed not only in big companies but also in individual houses. The evolution of the wireless local area network (WLAN) provides fast and convenient network connections. Live video streaming has been widely used nowadays, and the number of applications is continuously growing. For example, people use smartphones to watch video streaming for entertainment. In the field of security system, the installation of surveillance cameras can be flexible and economical if wireless networks are used to provide connections. During searching and rescuing operations, real-time audio and video communications in the wireless ad-hoc network can save lives.

Live streaming requires a steady stream of information and delivery of packets by a deadline. However, wireless networks have difficulties in providing such services in a reliable way, as the range of wireless home networks is typically limited and there is radio intermittent interference from external sources such as microwave ovens or cordless phones. For the mobile node, multi-path fading and shadow may further increase the link capacity and the variability of transmission error rate. The ability to transport multimedia content over a variety of networks at different channel conditions and bandwidth capacities with various requirements of quality-of-service (QoS), is considered to be a fundamental challenge for future communication systems. The end-to-end performance of the live video streaming service is effected jointly by video coding, reliable transport, and wireless resource allocation. Therefore, performance-aware adaptation techniques have become hot research topics to achieve optimal and dynamic network configurations at all times and for all network conditions.

The paper provides an overview of the video steaming service over wireless networks and presents a simulation platform with ns-3 network simulator. The structure of this paper is organized as follows. Section \ref{sec:background} describes the background and motivations of the project. Section \ref{sec:related-work} describes related works. Section \ref{sec:problem} discusses the problem that is studied in the project. Section \ref{sec:method} illustrates the proposed simulation platform. Section \ref{sec:results} presents the experimental results and analysis. Section \ref{sec:conclusion} concludes the paper.

\section{Background and Motivations}
\label{sec:background}

\subsection{Video Streaming}

Streaming is a technology used to deliver content from the server to clients over the internet without having to download it. Multimedia streaming is one of the most popular and successful streaming services since it allows the user to watch the video or listen to music almost immediately without having to wait for the file to be completely downloaded. Unlike the file transfer which keeps the file on the device until the user manually deletes it, the streaming data is automatically removed after the user uses it.

Video streaming requires a relatively fast internet connection. For services like Hulu, YouTube, and Netflix, 2-3 Mbps are required for SD, 5-8 Mbps for HD, and 12-25 Mbps for UHD. Live streaming uses the same techniques, but it is specifically designed for real-time internet content delivery. Live streaming is popular when it comes to viewing and interacting with live concert shows, gaming broadcasts, and special events or sports.

User Datagram Protocol (UDP) is preferred over Transmission Control Protocol (TCP) for video streaming. Unlike TCP, UDP does not send message back and forth to open a connection before transmitting data, and it does not ensure that all data packets arrive in order. As a result, transmitting data does not take as long as it does via TCP. Even though some packets are lost during the transmission, there are abundant data packets involved in keeping a video stream playing that the user would not notice the lost ones.

\subsection{Adaptive Video Streaming}

 Traditional progressive video streaming is simply one single video file being streamed over the internet, and the video can be stretched or shrunk to fit different screen resolutions. Regardless of the device playing it, the video file will always be the same. However, the technique brings two main problems: (1) The device with higher screen resolution than the video resolution would possibly encounter pixelation; (2) The device with poor internet connection, which cannot download the video stream quickly enough, will need to pause for receiving more video data to the buffer (this is also referred to as {\em rebuffering}). Either situation will bring a horrible video streaming experience for the end user. 
 
 Adaptive streaming (also known as {\em Adaptive Bitrate Streaming}), instead, is a technique designed to deliver the multimedia contents to the user in the most efficient way and in the highest possible quality for each user. Specifically, adaptive streaming needs the video streaming server to create a different video file for each target screen size, and it will lower the video quality for the device with slow internet connection.

\subsection{Wireless Network and Access Point}

A wireless network allows devices to connect to the internet without any cables. Access points amplify WiFi signals, so a device can be far from a router but  still be connected to the network. The wireless networks are usually realized and administered using the radio communications. Wireless networks such as Wireless Local Area Network (WLAN) and 4G go for high data rate applications, which will perfectly fit with the future requirements of the wireless video streaming applications.

An access point receives data by wired Ethernet, and converts to a 2.4GHz or 5GHz wireless signal. It sends and receives wireless traffic to and from nearby wireless clients. An access point is different from a wireless router in that it does not have firewall functions and does not protect your local network against threats from the Internet.

When you set up your access point using a wired connection, the access point functions as a WiFi base station or, if you use a mesh WiFi network, as a root access point. With a wired connection to the Internet, your access point can function as a WiFi base station or root access point for up to four other access points. These access points function as WiFi repeaters or, if you use a mesh WiFi network, extender access points. These access points connect to the Internet through the WiFi base station or root access point over a WiFi connection.

\subsection{ns-3 Network Simulator}

ns-3 \cite{ns-3doc} is a discrete-event network simulator for internet systems, and has been developed to provide an open, extensible network simulation platform, for networking research and education.

ns-3 provides models of how packet data networks work and perform, and provides a simulation engine for users to conduct simulation experiments. Some of the reasons to use ns-3 include to perform studies that are more difficult or not possible to perform with real systems, to study system behavior in a highly controlled, reproducible environment, and to learn about how networks work. Users will note that the available model set in ns-3 focuses on modeling how Internet protocols and networks work, but ns-3 is not limited to Internet systems; several users are using ns-3 to model non-Internet-based systems.

\section{Related Works}
\label{sec:related-work}

Video streaming is considered one of the most prevalent technologies, and has been studied in the past several years as the research topic. 

\cite{zhu2007video} discusses different wireless streaming scenarios, ranging from the simple case of delivering a single video stream over a single wireless link, to sharing a wireless multi-access channel among multiple video streams, to the general case of multiple streams sharing a mesh network. It also introduces two types of network topology for multiple video streams sharing a single-hop wireless network (See Section \ref{sec:problem} for details). By analyzing the results from the simulation, they validate the accuracy and performance of the model by computing the transmission rate and effective maximum throughput.

\cite{fouda2014real} presents a ns3-based real-time video streaming emulation platform for evaluating and optimizing the performance of the LTE networks. Three sample test cases are studied in the paper to verify the developed platform, which are video client mobility, streaming video to multiple clients and handover over the HIL platform. Moreover, it also evaluates multiple streaming protocols such as UDP, RTP, and Dynamic Adaptive Streaming over HTTP (DASH) with the emulation platform.

\cite{huang_buffer-based_2014} introduces a buffer-based approach to rate adaption which tries to solve the challenge in estimating future capacity. The method uses {\em only} the buffer to choose a video rate, and then ask {\em when} capcity estimation is needed. There are usually two separate phases of operation in video streaming service: a {\em steady-state} phase when the buffer has been built up, and a {\em startup} phase when the buffer is still growing from empty. The paper tests the viability of the approach through a series of experiments spanning millions of real users in a commercial service, and reveals that capacity estimation is unnecessary in steady state; however using simple capacity estimation is crucial during the startup phase.

\cite{liu_when_2020} proposes an integration of wireless multimedia systems and deep learning, which starts with decomposing a wireless multimedia system into three components, including end-users, network environment, and servers. Then the paper presents two cases, deep learning based QoS/QoE prediction and bitrate adjustment. In the former case, an end-to-end and unified DNN architecture was devised to fuse different types of multimedia data and predict the QoS/QoE value. In the latter case, a deep reinforcement learning based framework was designed for bitrate adjustment according to the viewer's interests. By evaluating the performance with a real wireless dataset, the deep learning approach can improve the video QoE average bitrate, rebuffering time, and bitrate variation significantly.

\section{Problem Definition and Thesis Statement}
\label{sec:problem}

\subsection{Problem Definition}

The project is to design a video streaming application with adaptive rate controller on top of UDP protocol. It is easy for a streaming service to meet either one of the objectives on its own. To maximize video quality, a service could just stream at the maximum video rate all the time. Of course, this would risk extensive rebuffering. On the other hand, to minimize rebuffering, the service could just stream at the minimum video rate all the time, which would lead to low video quality. The approach in our project is to dynamically change the video rate based on the link speed and frame buffer of the client, and thus ensure the best viewing experience for the user.

\subsection{Project Objectives}

\begin{figure}[t]
    \centering
    \includegraphics[scale=1.0]{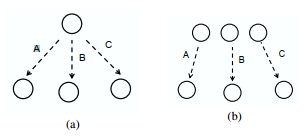}
    \caption{Two different network topology for multiple video streams sharing
a single-hop wireless network.}
    \label{fig:topology}
\end{figure}

The project is designed to provide a simulation of video streaming over wireless home networks in different scenarios, with increasing complexity. We first simulate the simple scenario of delivering a single video stream over a single wireless link, and then the case of sharing a wireless multi-access channel among multiple video streams. We plan to use different video files, such as 480p, 720p, 1080p, and so on, for different screen resolutions and network speed. The video streaming task is performed over simulated 802.11a home networks. The streaming application transmits video frames with UDP packets and contains an application-layer rate controller that can switch between different versions of video bitstreams based on the feedback from the client.

The project aims to simulate two kinds of network topology for multiple video streams sharing a single-hop wireless network \cite{zhu2007video} in Figure \ref{fig:topology}. 
\begin{itemize}
    \item All streams originate from the same wireless node. 
    \item The video source nodes are distributed. 
\end{itemize}

\section{Methodology}
\label{sec:method}

The simulation platform includes three main parts: video streaming server, video streaming client, and simulated point to point and Wirelss network.

\subsection{Video Streaming Server}

The main task of the video streaming server is to transmit the video data to the client(s). A maximum packet size is set for video data transmission, the number is 1400 bytes in our case. Nevertheless, the size of video frame can be greater than given maximum packet size, e.g., the size of one frame from high-definition is approximately 500 KB. Hence, the server is required to break the frame into several packets and deliver each of them to the client. 

Since the server is capable of delivering video files to multiple clients, it needs to handle connections from different clients and make sure that the video file is sent to corresponding client. Algorithm \ref{alg:handlerecv} demonstrates the behaviors of the server for handling the packets received from the client. The server first receives a socket from the client, and it will get the sender IP address from the socket. Then, the server will look up the hash table of the clients to check if the IP address already exists. If not, it means a new client tries to connect to the server, and the sever will add the new client to the hash table (line 4-9). Otherwise, the server will treat the packet as an request from the client to either raise or lower the video resolution (line 10-14).

\begin{algorithm}
\caption{Server handles packet receiving}
\label{alg:handlerecv}
\begin{algorithmic}[1]
\Function{HandleReceive}{}

\State $socket\gets ReceiveSocket ()$ 
\State $ipAddr \gets socket.GetSendingIPAddress ()$ 
\If {$clientTable.Find (ipAddr) == 0$} 
    \State $client = CreateNewClient ()$
    \State $client.sentNumber = 0$
    \State $client.videoLevel = 3$
    \State $client.ipAddress = ipAddr$
    \State $client.scheduleSendEvent ()$
\Else 
    \State $client = clientTable.Get(ipAddr)$
    \State $buffer \gets packet.ReadData()$
    \State $newLevel \gets buffer.ReadValue()$
    \State $client.UpdateLevel(newLevel)$

\EndIf
\EndFunction
\end{algorithmic}
\end{algorithm}

\subsection{Video Streaming Client}

The video streaming client is to receive the video frames from the server and display it. The client contains a playback buffer storing the video frames, and it reads the video frames from the buffer every second. If the buffer does not contain enough frames, the client will pause the video and replay it until it receives enough frames from the remote server. 

The rate controller is based on the size of playback buffer at the client side, and Algorithm \ref{alg:readbuffer} shows how the client deals with the playback buffer. First the client checks if the buffer has enough frames to play the video. If not, the client will decide if it needs to request the lower video quality (line 9-14) or stop the application (line 4-7) by comparing the current buffer size with the last recorded size. If the video can be played as expected, the client will also examine if the buffer stores much more frames than required for one second, then decide if it needs to request higher video quality. 

\begin{algorithm}
\caption{Client reads data from buffer}
\label{alg:readbuffer}
\begin{algorithmic}[1]
\Function{ReadBuffer}{}
\If {$curBufferSize < frameRate$}
    \If {$lastBufferSize == curBufferSize$}
        \State $stopCounter += 1$
        \If {$stopCounter \geq 3$}
            \State $StopClient()$
            \State \Return 
        \EndIf
    \Else
        \State $stopCounter = 0$
        \State $rebufferCounter += 1$
        \If {$rebufferCounter \geq 3$}
            \State $RequestLowerQuality()$
        \EndIf
    \EndIf
\Else
    \State $stopCounter = 0$
    \State $rebufferCounter = 0$
    \State $PlayFromBuffer()$
    \If {$curBufferSize > 5 * frameRate$}
        \State $RequestHigherQuality()$
    \EndIf
\EndIf

\State $ScheduleBufferEvent()$

\EndFunction
\end{algorithmic}
\end{algorithm}

% \subsection{Point-to-Point and Wireless Network}
\subsection{Wireless Network}

% Our point-to-point networks consist of one-to-one and one-to-multiple connections. After creating the point to point server nodes with port and client nodes, install the video streaming client and server app at the server and client nodes, then run the simulation. 

The wireless network used in the simulation follows IEEE 802.11 standard. Algorithm \ref{alg:wireless} shows the way we simulate the wireless network. Two helper classes \texttt{YansWifiChannelHelper} and \texttt{YansWifiPhyHelper} are used to set up \texttt{Channel} and \texttt{WifiPhy}. Default propagation delay model is \texttt{ConstantSpeedPropagationDelayModel}, and the default loss model is \texttt{LogDistancePropagationLossModel}. Default bit rate model: \texttt{NistErrorRateModel}, use the default \texttt{PHY} layer configuration and channel model. Then configure the Mac type and basic settings. To set up \texttt{WifiMac} and install \texttt{NetDevice} in the node, two helper classes \texttt{WifiHelper} and \texttt{WifiMacHelper} are used. The \texttt{SetRemoteStationManager} method tells the helper class which rate control algorithm to use, here is the AARF algorithm. Create an IEEE 802.11 service set identifier (SSID) object to set the "SSID" attribute value of the MAC layer. The MAC created by the helper class will not send a probe request. The setting will not send out the command to actively detect AP, the default AP is SSID. SSID is the abbreviation of Service Set Identifier. SSID technology can divide a wireless local area network into several sub-networks that require different authentication. Each sub-network requires independent authentication. Only users who pass the authentication can enter the corresponding sub-network. Network to prevent unauthorized users from entering the network.

Then set up the mobile model, we hope that the STA node is mobile, roaming within a bounding box, and we hope that the AP node remains stationary. ns-3 uses the Cartesian coordinate system to identify node positions, and the helper class is \texttt{MobilityHelper}. Set the mobile model for the mobile node. The mobile node model setting is divided into two parts: the initial position distribution and the subsequent movement trajectory model. Initial position distributor: \texttt{GridPositionAllocator}. The nodes will be equally spaced in a two-dimensional Cartesian coordinate system according to the set row and column parameters. Movement trajectory model: \texttt{RandomWalk2dMobilityModel}, the node moves at random speed in a rectangular area of specified size, the default value range is 2-4m/s and random direction. Then set the fixed position model for the AP node, using a fixed position mobile model \texttt{ConstantPositionMobilityModel}, the two-dimensional coordinates of the AP node of this model are (0, 0).

Then we install server and client programs, and set properties for them, e.g., \texttt{MaxFrame = 500, Interval = 0.01s, PacketSize = 1400 bytes}.

We visualized the network topology with PyViz. PyViz is a coordinated effort to make data visualization in Python easier to use, learn and more powerful. PyViz consists of a set of open-source Python packages to work effortlessly with both small and large datasets right in the web browsers.

\begin{algorithm}
\caption{Simulate wireless network}
\label{alg:wireless}
\begin{algorithmic}[1]
\Function{WirelessSimulation}{}

\State $wifiStaNodes = CreateNodes(numSta)$
\State $wifiApNodes = CreateNodes(numAp)$
\State $YansWifiChannelHelper\gets Default ()$ 
\State $YansWifiPhyHelper \gets Default ()$ 
\State $WifiHelper \gets AarfWifiManager()$
\State $WifiMacHelper \gets SetSsid()$
\State $wifi.Install (phy, mac, $
\State $wifiStaNodes, wifiApNodes)$

\State $wifiApNodes \gets SetMobilityModel() $

\State $setBaseAddress()$
\State $wifiApNodes.install(serverApps)$ 
\State $wifiStaNodes.install(clientApps)$
\State $PopulateRoutingTables ()$
\EndFunction
\end{algorithmic}
\end{algorithm}

\subsection{Testing Method}

\begin{figure*}[ht]
    \centering
    \begin{tabular}{c c}
    \includegraphics[height=3cm]{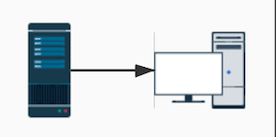} & \quad 
    \includegraphics[height=4cm]{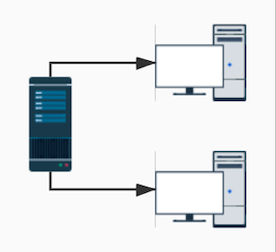} \\
    (a) & (b) \\
    \includegraphics[height=4cm]{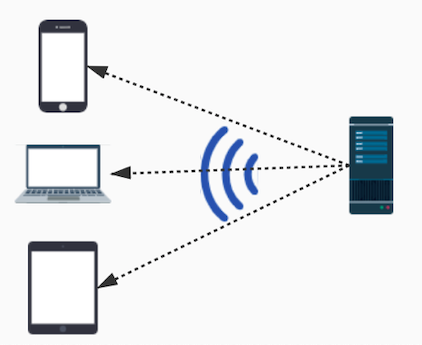} & \quad
    \includegraphics[height=4cm]{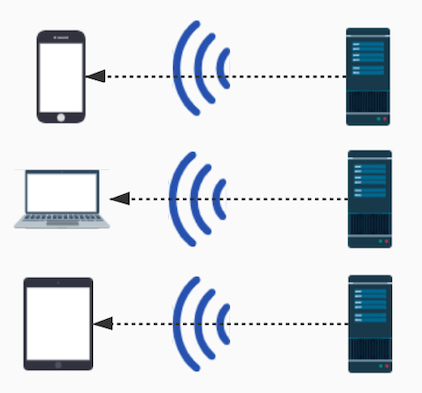} \\
    (c) & (d) \\
    \end{tabular}
    
    \centering
    \caption{Different network environments to be tested with the simulation, from wired P2P network to wireless home network.}
    \label{fig:paradigm}
\end{figure*}

To verify the developed simulation platform, we will set the link speed and the size of the video packet. By collecting the results from the simulation, we can validate the accuracy and performance of the model by computing the transmission rate and effective maximum throughput.

We will run our simulation on four scenarios, which are shown in Fig. \ref{fig:paradigm}.

\begin{enumerate}
    \item[(a)] Wired network with the point-to-point link between one server and one client.
    \item[(b)] Wired network with point-to-point connections between one server and two clients.
    \item[(c)] Wireless network with one access point (AP) server and one mobile client.
    \item[(d)] Wireless network with three AP servers and three mobile clients.
\end{enumerate}

The entire simulation will be performed in the virtual machine, and the guest OS installed on the VM is Ubuntu 18.04. The software that will be used in the project includes ns-3 (version 3.30), Python 3.6, Pyviz GUI.

\begin{figure*}[th]
    \centering
    \begin{tabular}{c c}
    \includegraphics[width=8cm]{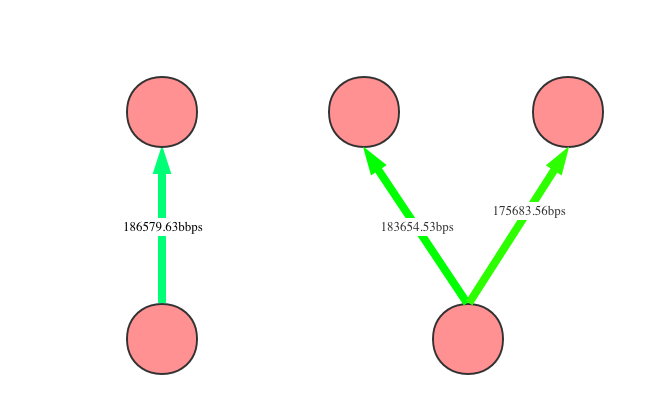} & 
    \includegraphics[width=8cm]{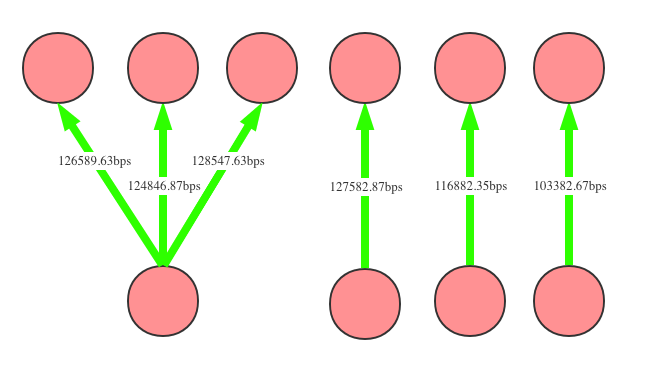} \\
    (a) & (b) \\
    \end{tabular}
    \caption{Simulation results for four scenarios aforementioned in the {\em Testing Method} section. (a) shows the application-level throughput in the wired environment. (b) shows the application-level throughput in the wireless environment.}
    \label{fig:results}
\end{figure*}

\section{Results}
\label{sec:results}

\subsection{Wireless Network Topology}

We simulated all scenarios mentioned previously, and the visualization is shown as Fig. \ref{fig:results}. We observe comparable application-level throughput in four different network environments if the internet speed is configured in similar way. 

\subsection{Video Stream Server}

In the experiment, we use the text file that stores a sequence of video frame sizes to pretend the video file. A sample text file is given as follows. The number in each line denote a frame size in bytes. The first line in the sample text means the first frame of the video is 22500 bytes.

\begin{minted}[frame=lines, breaklines, fontsize=\scriptsize]{text}
22500
1027
1027
1251
\end{minted}

The video stream server is able to break the large frame into several packets and deliver them to the client. The below shows the output for the first frame in the sample text file.
\begin{minted}[frame=lines, breaklines, fontsize=\scriptsize]{text}
At time 1s server sent 1400 bytes to 10.1.1.2 port 49153
At time 1s server sent 1400 bytes to 10.1.1.2 port 49153
At time 1s server sent 1400 bytes to 10.1.1.2 port 49153
At time 1s server sent 1400 bytes to 10.1.1.2 port 49153
At time 1s server sent 1400 bytes to 10.1.1.2 port 49153
At time 1s server sent 1400 bytes to 10.1.1.2 port 49153
At time 1s server sent 1400 bytes to 10.1.1.2 port 49153
At time 1s server sent 1400 bytes to 10.1.1.2 port 49153
At time 1s server sent 1400 bytes to 10.1.1.2 port 49153
At time 1s server sent 1400 bytes to 10.1.1.2 port 49153
At time 1s server sent 1400 bytes to 10.1.1.2 port 49153
At time 1s server sent 1400 bytes to 10.1.1.2 port 49153
At time 1s server sent 1400 bytes to 10.1.1.2 port 49153
At time 1s server sent 1400 bytes to 10.1.1.2 port 49153
At time 1s server sent 1400 bytes to 10.1.1.2 port 49153
At time 1s server sent 1400 bytes to 10.1.1.2 port 49153
At time 1s server sent 100 bytes to 10.1.1.2 port 49153
\end{minted}

\subsection{Video Stream Client}

The video stream client is required to monitor the status of the playback buffer, and handle different conditions of the buffer. 

\begin{itemize}
    \item The case of requesting a lower video quality: 
\begin{minted}[frame=lines, breaklines, fontsize=\scriptsize]{text}
(......)
At time 3.5 s: Not enough frames in the buffer, rebuffering!
At time 3.61909s client received frame 15 and 56408 bytes from 10.1.1.1 port 5000
At time 3.85341s client received frame 16 and 57350 bytes from 10.1.1.1 port 5000
At time 4.08793s client received frame 17 and 57400 bytes from 10.1.1.1 port 5000
At time 4.33389s client received frame 18 and 60200 bytes from 10.1.1.1 port 5000
At time 4.5 s: Not enough frames in the buffer, rebuffering!
At time 4.58276s client received frame 19 and 60898 bytes from 10.1.1.1 port 5000
At time 4.7944s client received frame 20 and 51800 bytes from 10.1.1.1 port 5000
At time 5.05752s client received frame 21 and 64400 bytes from 10.1.1.1 port 5000
At time 5.27239s client received frame 22 and 52577 bytes from 10.1.1.1 port 5000
At time 5.5 s: Not enough frames in the buffer, rebuffering!
At time 5.50119s: Lower the video quality level!
(......)
\end{minted}
    
    \item The case of requesting a higher video quality:
\begin{minted}[frame=lines, breaklines, fontsize=\scriptsize]{text}
(......)
At time 2.5349s client received frame 120 and 277397 bytes from 10.1.1.1 port 5000
At time 2.55769s client received frame 121 and 278819 bytes from 10.1.1.1 port 5000
At time 2.58043s client received frame 122 and 278288 bytes from 10.1.1.1 port 5000
At time 2.60336s client received frame 123 and 280667 bytes from 10.1.1.1 port 5000
At time 2.61411s client received frame 124 and 131446 bytes from 10.1.1.1 port 5000
At time 2.61411s: Increase the video quality level to 4
(......)
At time 4.11853s client received frame 245 and 120400 bytes from 10.1.1.1 port 5000
At time 4.1286s client received frame 246 and 123200 bytes from 10.1.1.1 port 5000
At time 4.13844s client received frame 247 and 120400 bytes from 10.1.1.1 port 5000
At time 4.14839s client received frame 248 and 121800 bytes from 10.1.1.1 port 5000
At time 4.15835s client received frame 249 and 121800 bytes from 10.1.1.1 port 5000
At time 4.16841s client received frame 250 and 123200 bytes from 10.1.1.1 port 5000
At time 4.16841s: Increase the video quality level to 5

\end{minted}
    
\end{itemize}

\section{Conclusion}
\label{sec:conclusion}

In this paper, we give an overview of the video streaming service and wireless network. The combination of two techniques can be applied in a variety of fields. We implemented an adaptive video streaming application in ns-3 simulator, which can deliver the video at different bitrates based on the internet speeds of connected clients. We test our implementation in various network scenarios, from a simple P2P network with only one client, to wireless networks with multiple servers and multiple clients. The results validate the accuracy and efficiency of our adaptive video streaming application and wireless network simulation.

We noticed that the mobile devices will lose connection to the server when it moves out of the range of wireless signals. However, we did not observe the transmission rate dropping when the mobile devices are away from the sever, which requires further investigation on how different modes work in the ns-3 wireless simulator. It is also worth noting that the simulation is under the most ideal circumstances, without any obstacle, electromagnetic interference, or air loss, which could possibly make our simulation results differ from real life scenarios. Therefore, future research should be conducted in more realistic settings to produce more realistic results.

\bibliography{ref}
\bibliographystyle{ieeetr}

\end{document}